\documentstyle[12pt,aps,subeqnarray,cite,verbatim]{revtex}
\begin{document}
\newcommand{\dis}{\displaystyle}
\newcommand{\dml}{\rm dim ~ }
\newcommand{\krl}{\rm ker ~ ~ }
\newcommand{\expon}{\rm e}
\newcommand{\id}{ 1 \hspace{-2.85pt} {\rm I} \hspace{2.5mm}}
\newcommand{\func}{ (\frac{\Phi(\Gamma) - \Phi(J_0)}{\Gamma -
[J_0][J_0 + 1]})^{\frac{1}{2}} }
\newcommand{\D}{\Delta}
\newcommand{\nJ}{\tilde{J}}
\newcommand{\ca}{{\cal C}}
\newcommand{\U}{{\cal U}}
\newcommand{\alp}{\alpha^\prime}
\newcommand{\beq}{\begin{equation}}
\newcommand{\eeq}{\end{equation}}
\newcommand{\bseq}{\begin{subeqnarray}}
\newcommand{\eseq}{\end{subeqnarray}}
\renewcommand{\thefootnote}{\fnsymbol{footnote}}
\newcommand{\Jp}{\tilde{J}_+}
\newcommand{\Jm}{\tilde{J}_-}
\newcommand{\Jz}{\tilde{J}_0}
\newcommand{\Jpm}{\tilde{J}_{\pm}}
\title{On Bures fidelity of displaced squeezed thermal states}

\author{Xiang-Bin Wang 
{\footnote{E-mail address:
scip7236@leonis.nus.edu.sg }},
C. H. Oh 
{\footnote{E-mail address:
phyohch@leonis.nus.edu.sg }} and
L. C. Kwek 
{\footnote{E-mail address:
scip6051@leonis.nus.edu.sg }}
}
\address{Department of Physics, Faculty of Science, \\
National University of Singapore, Lower Kent Ridge, \\
Singapore 119260, Republic of Singapore. }

\maketitle


\begin{abstract}
Fidelity plays a key role in quantum information and communication
theory. Fidelity can be interpreted as the probability that a decoded
message possesses the same information content as the message prior to
coding and transmission. In this paper, we 
give a formula of Bures fidelity for displaced squeezed thermal
states directly by the displacement and squeezing parameters and
birefly discuss how the results can apply to quantum information
theory.  
\end{abstract}


An important tenet in classical information theory is the rigorous
establishment of the Shannon Noiseless Coding Theorem, in which one
shows that Shannon entropy can be interpreted as the average number of
bits needed to code the output of a message source under ideal
conditions. The analogous quantum version of Shannon's Coding Theorem
is the Schumacher Quantum Coding Theorem \cite{schumacher}. In the
quantum version, one introduces the idea of fidelity which can 
be interpreted as the probability that a decoded message carries the
same information as the message prior to coding and replaces Shannon
entropy with fidelity. More specifically, one can proves the Schumacher
Noiseless Coding Theorem which states that if $M$ is a quantum signal
source with signal ensemble described by the density operator $\rho$
then $\forall ~ ~ \delta, \epsilon > 0$, 
\begin{itemize}
\item if $S(\rho) + \delta$ qubits are available per $M$ signal, then,
for sufficiently large $N$, groups of $N$ signal from the signal source
$M$ can be transposed through the available qubits with fidelity $F > 1
- \epsilon$.
\item if $S(\rho) - \delta$ qubits are available per $M$ signal, then,
for sufficiently large $N$, groups of $N$ signals from the signal
source $A$ can be transposed through the avaiable qubits with fidelity
$F < \epsilon$.
\end{itemize}

Suppose a quantum signal source, $M$, generates a signal state $|i_A>$
with probability $p(a)$ and the density operator, $\rho$, is described by 
the equation
\beq
\rho = \sum_a p(a) |a_M><a_M|,
\eeq
one can define the Schumacher fidelity, $F$, as the overall
probability that a signal from an ensemble $M$ can be transmitted to
$M^\prime$ using the relation \cite{schumacher, barnum}
\beq
F = \sum_{a} p_i {\rm Tr} (\pi_a \rho_a^\prime),
\eeq
where $p_i \equiv |a_M><a_M|$ and $\rho_a^\prime$ denotes the density operator
of the final signal in $M^\prime$.  This definition applies strictly
to pure states and it is generally not clear how it can be applied to
mixed states.

Closely related to the problem of coding is the process of entanglement
purification protocol (EPP) and quantum error-correction codes (QECC)
\cite{bennett1, bennett2}.  These
protocols essentially shield quantum states from the environment.  In
EPP, maximally entangled states are extracted (or purified) from a
mixed states while in QECC, an arbitrary quantum state is transmitted
at some rate through a noisy channel with miminmal  degradation.
Central to the idea of entanglement is the need to define a measure of
entanglement. Bennett and others have proposed a measure of
entanglement using the von Neumann entropy.  However, it is
sometimes difficult to compute and obtain a closed form using their
definition. Recently, Vedral and others have studied a wdie class of 
measures suitable for
entanglement and they have proposed the Bures metric as an
example of a possible means of quantifying entanglement or fidelity
\cite{vedral}.

Finally, we note that experimentally, 
it is well-known that a squeezed electromagnetic field
\cite{garrett} 
provides a means to overcome the standard quantum limit for noise
imposed by vacuum fluctuations.  Furthermore, although the number-state
channel is an optimal channel for quantum communication theory, it is
often more realistic to consider the quadrature-squeezed channel
\cite{ariano} experimentally for several reasons.  Firstly, one cannot
faithfully reproduce the number eigenstates easily and secondly
amplification of quadrature-squeezed channel can be realized
experimentally using a phase-sensitive amplifier. Clearly, one should
therefore investigate the plausibility of applying squeezed or
displaced squeezed thermal states to quantum information and 
communication theory.

Indeed, Bures fidelity has been an important concept in the field of quantum
optics. Recently, Twamley \cite{twamley} has calculated the Bures fidelity for
squeezed thermal states. Due to some technical difficulties, the
displaced squeezed states was not considered in his article. Very recently,
Scutario \cite{scutaru} proposed an approach  to calculate the Bures
fidelity for systems with quadratic Hamiltonian. However, a closed form
for the matrix elements of the density operator is not explicitly given
and the final result does not relate fidelity directly with the squeezing
and displacement parameters. However, in a more recent paper
\cite{scutaru2}, he has obtained an explicit form for the Bures
fidelity for two displaced 
thermal states. In this article, we show an alternative
method in which one can actually calculate the fidelity of displaced squeezed
thermal states by simply using the Baker-Compell-Hausdorf(BCH) formula.
A closed form result for displaced squeezed thermal states expressed in
squeezing and displacement parameters is also obtained. 

Squeezed states occur in a myriad of non-linear optical phenomena like
optical parametric oscillation and four-wave mixing \cite{barnett}. 
The single-mode
squeezed states can be generated from the vaccuum by the action of the
squeezed operator $S$,
\beq
S(\zeta) = \exp(\frac{1}{2}(\zeta^\ast a^2 - \zeta a^{\dagger 2})),
\eeq
where $\dis \zeta = r \expon^{i \phi}$ is a complex number with modulus
$r$ and argument $\phi$, representing the squeezing parameter.
The density
operator of displaced squeezed thermal states can be defined as:
\begin{eqnarray} 
\rho=DS\Lambda S^{+}D^{+}
\end{eqnarray} 
where $D=\exp\left[(a^{+},a)\left(\begin{array}{c}
k\\-k^{*}\end{array}\right)\right]$,
$S=\exp\left[\frac{1}{2}r\left((a^{2}-a^{+2}\right)\right)$ and
$\Lambda = \exp[-\frac{1}{2}(aa^{+}+a^{+}a)]$. Note that we have
considered the squeezing parameter to be real since the most important
parameter of a squeezed state is the squeezed factor $r$ and not its
argument $\phi$\cite{schumaker}.  The general case in
which the argument $\phi$ is nonzero can be treated similarly.
We next recall that the Bures
fidelity, $F$, can be defined by the relation
\begin{eqnarray} F =  
\left({\rm
tr}\sqrt{{\rho_1}^{\frac{1}{2}}\rho_2{\rho_1}^{\frac{1}{2}}}\right)^2.
\end{eqnarray} 
For two displaced
squeezed thermal states, one easily sees that the Bures fidelity can be
expressed as
\bseq
F  & = &   \left( {\rm tr } \sqrt{(D_1 S_1 \Lambda_1^{1/2} S_1^\dagger
D_1^\dagger)(D_2 S_2 \Lambda_2 S_2^\dagger
D_2^\dagger)(D_1 S_1 \Lambda_1^{1/2} S_1^\dagger
D_1^\dagger)  }\right)^2.  \\
& = & \left( {\rm tr } \sqrt{ \Lambda_1^{1/2} S_1^\dagger
D_1^\dagger D_2 S_2 \Lambda_2 S_2^\dagger
D_2^\dagger D_1 S_1 \Lambda_1^{1/2}   }\right)^2
\label{bures}
\eseq
To simplify eq(\ref{bures})\cite{wang}, we need to rewrite ${D_1}^{+}D_2$ as 
\bseq{D_1}^{+}D_2  =  c \cdot D_0=c \cdot
\exp\left[ (a^{+},a)\left( \begin{array}{c}
g \\
-g^{*}\end{array} \right) \right] 
\mbox{{\rm and ~ ~ ~}} D_2^\dagger D_1 & = & \frac{1}{c} \cdot
D_0^\dagger \label{eqn3}
\eseq
where
$\dis 
\left(\begin{array}{c}
g\\-g^{*}\end{array}\right)=\left(\begin{array}{c}
k_2-{k_1}^*\\-\left(k_2-{k_1}^*\right)^{*}\end{array}\right)\
$ and $c$ is a number.  Thus, one can rewrite
the formula for Bures fidelity of displaced
squeezed thermal states appears as
\begin{eqnarray}F= \left({\rm tr}\sqrt{{\Lambda_1}^{\frac{1}{2}}
{S_1}^{+}{D_0}^{+}S_2{\Lambda}_2D_0S_1{\Lambda_1}^{\frac{1}{2}}}\right)^2
\label{fid}
\end{eqnarray} 
with $D_0 ={D_1}^{+}D_2$.
Eq(\ref{fid}) needs some simplication before we can actually proceed
with the detailed calculations.  Before we do this, we need to invoke the 
BCH relation,
\begin{eqnarray}
S(a^{+},a)S^{+}=(a^{+},a)M; ~ ~ S^{+}(a^{+},a)S=(a^{+},a)M^{-1}; 
\end{eqnarray}
where $M=\left(\begin{array}{cc}{\rm ch} r & -{\rm sh}r\\-{\rm sh}r &
{\rm ch}r \end{array} \right)$ and
\beq
\Lambda (a^{+},a) \Lambda^{-1}=(a^{+},a) B. \label{bogo}
\eeq
Note that in eq(\ref{bogo}), we have introduced the matrix 
$B \equiv \left(\begin{array}{cc}\exp (-\beta )  & 0\\ 0
& \exp ( \beta ) \end{array} \right)$.  
Let us define the matrix $\Omega$ as
${\Lambda_1}^{\frac{1}{2}}
{S_1}^{+}{D_0}^{+}S_2{\Lambda}_2 S_2^\dagger 
D_0S_1{\Lambda_1}^{\frac{1}{2}} $ in 
eq(\ref{fid}). It is instructive to note that, 
by using BCH formula, one can readily express the matrix $\Omega$ in a more
convenient form as
$$\Omega={\Lambda_1}^{\frac{1}{2}}
{S_1}^{+}S_2{{\Lambda}_2}^{\frac{1}{2}}
\exp\left[(a^{+},a){B_{2}}^{-\frac{1}{2}}{M_2}^{-1}
\left(\begin{array}{c}g\\-g{*}\end{array}\right)\right]$$
\begin{eqnarray}\times
\exp\left[-(a^{+},a){B_{2}}^{\frac{1}{2}}{M_2}^{-1}
\left(\begin{array}{c}g\\-g{*}\end{array}\right)
\right]{{\Lambda}_2}^{\frac{1}{2}} 
S_2^\dagger S_1{\Lambda_1}^{\frac{1}{2}} \label{eqn7}
\end{eqnarray}
The linear terms within the exponential factor in the abovee 
formula (\ref{eqn7}) 
can be collapsed into a simpler term by using the following results
(see Appendix A for a detailed proof): 
$$
\exp\left[(a^{+},a) N_1\left(
\begin{array}{c}z_1\\z_2
\end{array}\right)\right]\exp\left[(a^{+},a)N_2
\left(\begin{array}{c}z_3\\z_4\end{array}\right)\right]$$
\begin{eqnarray}=\exp\left[-\frac{1}{2} (z_1,z_2)\widetilde{N_1}\Sigma
N_2\left(\begin{array}{c}z_3\\z_4
\end{array}\right)\right]\exp\left[(a^{+},a)N_1
\left(\begin{array}{c}z_1\\z_2\end{array}\right)+(a^{+},a)N_2
\left(\begin{array}{c}z_3\\z_4\end{array}\right)\right] \label{append}
\end{eqnarray} 
where $N_1,N_2$ are arbitrary $2\times 2$ complex matrices, $\Sigma$ is
the matrix $\dis \left( \begin{array}{cc} 
0 & 1 \\
-1 & 0
\end{array} \right)$
and $z$ is arbitrary complex number. 
In this mannner, one sees that
\begin{eqnarray}
\Omega=\delta_1 \rho_+ 
\exp\left[(a^{+},a)({B_2}^{-\frac{1}{2}}-{B_2}^{\frac{1}{2}})
{M_2}^{-1}\left(\begin{array}{c}g\\-g^{*}\end{array}\right)\right]\rho_{-}
\end{eqnarray}
and 
\begin{eqnarray}
\delta_1 =\exp\left[\frac{1}{2} (g,-g^{*})
\widetilde{M_2}^{-1}{B_2}^{-\frac{1}{2}}\Sigma 
{B_{2}}^{\frac{1}{2}}{M_2}^{-1}\left(
\begin{array}{c}g\\-g{*}\end{array}\right)\right]
\end{eqnarray}
 $\rho_+ ={\Lambda_1}^{\frac{1}{2}} {S_1}^{+}S_2{{\Lambda}_2}^{\frac{1}{2}}$, 
$\rho_-={{\Lambda}_2}^{\frac{1}{2}}S_2^\dagger S_1{\Lambda_1}^{\frac{1}{2}}$\\
Let us now consider another operator
\begin{eqnarray}
\Omega'=U\rho_+\rho_-U^{+}
\end{eqnarray}
where $U=\exp\left[(a^{+},a)\left(\begin{array}{c}l\\-l^{*}\end{array}
\right)\right]$.
If we apply BCH formula again,  we shall see that
$$\Omega'=\rho_{+}\exp\left[(a^{+},a){B_2}^{-\frac{1}{2}}
{M_2}^{-1}M_1{B_1}^{-\frac{1}{2}}\left(
\begin{array}{c}l\\-l^{*}\end{array}\right)\right]$$
\begin{eqnarray}
\times 
\exp\left[-(a^{+},a){B_2}^{\frac{1}{2}}{M_2}^{-1}
M_1{B_1}^{\frac{1}{2}}\left(\begin{array}{c}l\\
-l^{*}\end{array}\right)\right]\rho_-
\end{eqnarray}
\begin{eqnarray}
\Rightarrow \Omega'=\delta_2\rho_{+}
\exp\left[(a^{+},a)({B_2}^{-\frac{1}{2}}{M_2}^{-1}M_1{B_1}^{-\frac{1}{2}}
-{B_2}^{\frac{1}{2}}{M_2}^{-1}M_1{B_1}^{\frac{1}{2}})
\left(\begin{array}{c}l\\-l^{*}\end{array}\right)\right]\rho_-
\end{eqnarray}
and
\begin{eqnarray}
\delta_2=\exp\left[\frac{1}{2} 
(l,-l^{*}){B_1}^{-\frac{1}{2}}\widetilde{M_1}
\widetilde{M_2}^{-1}{B_2}^{-\frac{1}{2}}\Sigma
{B_2}^{\frac{1}{2}}{M_2}^{-1}M_1{B_1}^{\frac{1}{2}}
\left(\begin{array}{c}l\\-l^{*}
\end{array} \right)\right] \label{delta2}
\end{eqnarray}
Setting
\begin{eqnarray}
({B_2}^{-\frac{1}{2}}{M_2}^{-1}
M_1{B_1}^{-\frac{1}{2}}-{B_2}^{\frac{1}{2}}
{M_2}^{-1}M_1{B_1}^{\frac{1}{2}})\left(
\begin{array}{c}l\\-l^{*}\end{array}\right)=(
{B_2}^{-\frac{1}{2}}-{B_2}^{\frac{1}{2}}){M_2}^{-1}\left(\begin{array}
{c}g\\-g^{*}\end{array}\right) \label{eqn19}
\end{eqnarray}
we get
\begin{eqnarray}
\Omega=\frac{\delta_1}{\delta_2}\Omega'
\end{eqnarray}
\begin{eqnarray}
\left({\rm tr}\sqrt{\Omega}\right)^{2}=
\frac{\delta_1}{\delta_2}\left({\rm tr}\sqrt{U\rho_+\rho_-U^+}\right)^{2}
=\frac{\delta_1}{\delta_2}\left({\rm tr}\sqrt{\rho_+\rho_-}\right)^{2}
\label{dist} 
\end{eqnarray}
Since $\left({\rm tr}
\sqrt{\rho_+\rho_-}\right)^{2}$ has already been computated 
in Ref\cite{twamley}, we can solved the whole problem by considering the
reduced calculation of $\frac{\delta_1}{\delta_2}$. Following Twamley
paper, one notes that the quantity $({\rm tr} \sqrt{\rho_+ \rho_-})^2 $ in
eq(\ref{dist}) can be written as
\beq
({\rm tr} \sqrt{\rho_+ \rho_-})^2 = \frac{2 \sinh \frac{\beta_1}{4}
\sinh \frac{\beta_2}{4}}{\sqrt{\sqrt{Y} - 1}}
\eeq
where $Y = \cosh^2 (r_1 - r_2) \cosh^2 \frac{\beta_1 + \beta_2}{4}
+ \cosh^2 (r_1 + r_2) \cosh^2 \frac{\beta_1 + \beta_2}{4} -
\sinh^2 (r_1 - r_2) \cosh^2 \frac{\beta_1 - \beta_2}{4} -
\cosh^2 (r_1 + r_2) \cosh^2 \frac{\beta_1 - \beta_2}{4}.
$
From eqs (\ref{delta2}) and (\ref{eqn19}), one quickly get:
$$\delta_2=\exp\left\{(l,-l^{*}){B_1}^{-\frac{1}{2}}\widetilde{M_1}
\widetilde{M_2}^{-1}{B_2}^{-\frac{1}{2}}\cdot\Sigma \cdot \right.$$
\begin{eqnarray}\times \left.
 \left[ {B_2}^{-\frac{1}{2}}{M_2}^{-1}M_1{B_1}^{-\frac{1}{2}}
\left(\begin{array}{c}l\\-l^{*}
\end{array}\right)-({B_2}^{-\frac{1}{2}}-{B_2}^{\frac{1}{2}})
{M_2}^{-1}\left(\begin{array}
{c}g\\-g^{*}\end{array}\right)\right]\right\}
\end{eqnarray}
It is instructive to note that 
the matrix B and M are all sympletic matrices, so that we have
\begin{eqnarray}
(l,-l^{*}){B_1}^{-\frac{1}{2}}\widetilde{M_1}
\widetilde{M_2}^{-1}{B_2}^{-\frac{1}{2}}
\Sigma {B_2}^{-\frac{1}{2}}{M_2}^{-1}M_1{B_1}^{-\frac{1}{2}}
\left(\begin{array}{c}l\\-l^{*}\end{array}\right)=
(l,-l^{*})\Sigma\left(\begin{array}{c}l\\-l^{*}
\end{array}\right)=0. \label{eqn23}
\end{eqnarray}
With this observation, it is straightforward to see that 
eq(\ref{delta2}) can be simplified as
\begin{eqnarray}
\delta_2=\exp\left[-(l,-l^{*}){B_1}^{-\frac{1}{2}}\widetilde{M_1}
\widetilde{M_2}^{-1}{B_2}^{-\frac{1}{2}}
\Sigma({B_2}^{-\frac{1}{2}}-{B_2}^{\frac{1}{2}}){M_2}^{-1}\left(\begin{array}
{c}g\\-g^{*}\end{array}\right)\right].
\end{eqnarray}
To obtain the final explicit form of $\delta_2$, one notes that 
from eq(\ref{eqn19}), we have
\begin{eqnarray}
(l,-l^{*})=(g,-g^{*})\widetilde{M_2}^{-1}
({B_2}^{-\frac{1}{2}}-{B_2}^{\frac{1}{2}})
\widetilde{P}^{-1}
\label{eqn14}
\end{eqnarray}
where the matrix $\dis P \equiv ({B_2}^{-\frac{1}{2}}{M_2}^{-1}
M_1B_1-{B_2}^{\frac{1}{2}}{M_2}^{-1}M_1{B_1}^{\frac{1}{2}})$.
If we plug eq(\ref{eqn14}) into the eq(\ref{eqn23}), 
we arrive at the following formula for calculation of $\delta_2$
$$\delta_2=\exp\left[-\frac{1}{2}(g,-g^{*})\widetilde{M_2}^{-1}
({B_2}^{-\frac{1}{2}}-{B_2}^{\frac{1}{2}})
\widetilde{P}^{-1}\right.$$
\begin{eqnarray}
\times \left. {B_1}^{-\frac{1}{2}}\widetilde{M_1}
\widetilde{M_2}^{-1}{B_2}^{-\frac{1}{2}}
\Sigma({B_2}^{-\frac{1}{2}}-{B_2}^{\frac{1}{2}}){M_2}^{-1}
\left(\begin{array}{c}g\\-g^{*}\end{array}\right)\right]
\end{eqnarray}

In our case, it is not difficult to evaluate the expression for
$\delta_1$ and $\delta_2$ explicitly.  To do this, we note that if we
denote 
\beq
\delta_1 = \exp \left[ \frac{1}{2}(g, - g^\ast) Q_1 
\left(\begin{array}{c}g\\-g^{*}\end{array}\right)
\right],
\eeq
then the matrix $Q_1$ is simply
\beq
Q_1 = \left(
\begin{array}{ll}
\sinh \beta_2 \sinh(2 r_2) & \cosh \beta_2  +  \sinh \beta_2 \cosh(2
r_2) \\
-\cosh \beta_2 + \sinh \beta_2 \cosh(2
r_2) & \sinh \beta_2 \sinh(2 r_2) 
\end{array} \right).
\eeq
For $\delta_2$, one notes that a straightforward 
computation for the matrix $P$ yields
\beq
P = \frac{1}{\Delta} \left( 
\begin{array}{ll}
\sinh \frac{\beta_2 + \beta_1}{2} \cosh(r_1 - r_2) &
\sinh \frac{\beta_2 - \beta_1}{2} \sinh(r_1 - r_2) \\
- \sinh \frac{\beta_2 - \beta_1}{2} \sinh(r_1 - r_2) &
- \sinh \frac{\beta_2 + \beta_1}{2} \cosh(r_1 - r_2) 
\end{array} \right)
\eeq
with $\Delta = \cosh \beta_1 \cosh \beta_2 + \sinh \beta_1 \sinh \beta_2
\cosh 2(r_1 - r_2) - 1$, so that if we denote
\beq
\frac{\delta_1}{\delta_2} = \exp\left\{ \frac{1}{2} (g , - g^\ast) R
\left(\begin{array}{c}g\\-g^{*}\end{array}\right) \right\},
\eeq
then a straightforward, albeit tedious, calculation yields
\begin{eqnarray}
R & = & 
\left( \begin{array}{cc}
0 & 1 \\
-1 & 0
\end{array} \right)
+ \frac{2 }{\Delta}  \sinh \beta_1 \sinh^2 \frac{\beta_2}{2} 
\left( \begin{array}{cc}
\sinh(2 r_1) & \cosh(2 r_1) \\
\cosh(2 r_1) & \sinh(2 r_1)
\end{array} \right) \nonumber \\
& & 
\mbox{\hspace{5mm}} + \frac{2 }{\Delta}
\sinh^2 \frac{\beta_1}{2} \sinh \beta_2 
\left( \begin{array}{cc}
\sinh(2 r_2) & \cosh(2 r_2) \\
\cosh(2 r_2) & \sinh(2 r_2)
\end{array} \right)
\end{eqnarray}
so that the factor $\dis \frac{\delta_1}{\delta_2}$ works out
explicitly into
\beq
\exp \left\{ 
\frac{1}{\Delta}
\left(\epsilon_1 + \epsilon_2
\right)
\right\} 
\eeq
where
\bseq
\epsilon_1 & = & \sinh \beta_1 \sinh^2 \frac{\beta_2}{2} 
\left[ \left( g^2 + g^{\ast 2} \right) \sinh 2r_1 
- 2 |g|^2 \cosh 2 r_1 \right] \\
\epsilon_2 & = &  \sinh^2 \frac{\beta_1}{2} \sinh \beta_2 
\left[ \left( g^2 +  g^{\ast 2} \right) \sinh 2
r_2 - 2 |g|^2 \cosh 2 r_2 \right]. 
\eseq 

One can easily show that that $\frac{\delta_1}{\delta_2} < 1$ as it should be
and that in the limit $g = g^\ast = 0$, the ratio reduces to unity so
that one obtains the Bures fidelity for the undisplaced squeezed states
as shown in ref \cite{twamley}.  Further, one should also note that in
the limit when $r \rightarrow 0$, one gets the Bures fidelity for
the displaced thermal coherent states.  This Bures fidelity is the same as  
the result previously obtained by Scutaru
\cite{scutaru2}. 


\appendix
\section{}
In this appendix, we shall explicitly show the proof for
eq(\ref{append}). For simplicity and convenience, we shall define
$\Omega_i$  as the expression
\beq
\Omega_i  =  (a^\dagger, a) N_i \left( 
\begin{array}{c}
z_{2 i - 1} \\
z_{2 i}
\end{array}
\right), ~ ~ \mbox{{\rm for ~ } i= 1,2}. 
\eeq
To show eq(\ref{append}), we need to compute $\expon^{\Omega_1}
\expon^{\Omega_2}$.
Since $N_1$ and $N_2$ are simply two arbitrary $2 \times 2$ matrices,
one can always write in all generality
\beq
N_1 = \left( 
\begin{array}{cc}
a & d \\
b & c
\end{array} \right), ~ ~
N_2 = \left( 
\begin{array}{cc}
e & h \\
f & g
\end{array}
\right)
\eeq
We next compute the commutator for $\Omega_1$ and $\Omega_2$.
\beq
\lbrack \Omega_1, \Omega_2 \rbrack = 
- (a z_1 + d z_2) (f z_3 + g z_4) + (b z_1 + c z_2) (e z_3 + h z_4).
\eeq
On the other hand, we should note that
\bseq
(z_1, z_2 ) \widetilde{N_1} \Sigma N_2 \left( 
\begin{array}{c}
z_3 \\
z_4
\end{array} \right)
& = & (z_1, z_2) \left( 
\begin{array}{cc}
a & b \\
d & c
\end{array} \right) 
\left( 
\begin{array}{cc}
0 & 1 \\
-1 & 0
\end{array} \right)
\left( 
\begin{array}{cc}
e & h \\
f & g
\end{array} \right)
\left( 
\begin{array}{c}
z_3 \\
z_4
\end{array} \right) \nonumber \\
& = & (a z_1 + d z_2, b z_1 + c z_2) 
\left( 
\begin{array}{cc}
0 & 1 \\
-1 & 0
\end{array} \right)
\left( 
\begin{array}{c}
e z_3 + h z_4 \\
f z_3 + g z_4
\end{array} \right) \nonumber \\
& = & 
(- b z_1 - c z_2, a z_1 + d z_2)
\left( 
\begin{array}{c}
e z_3 + h z_4 \\
f z_3 + g z_4
\end{array} \right) \\
& = & -(b z_1 + c z_2)(e z_3 + h z_4) + (a z_1 + d z_2)(f z_3 + g z_4) 
\eseq
Consequently, using Baker-Campbell-Hausdorff formula, on gets
\bseq
\expon^{\Omega_1} \expon^{\Omega_2} & = & \expon^{\frac{1}{2}
[\Omega_1, \Omega_2]} \expon^{\Omega_1 + \Omega_2} \\
& = & \exp \left[- \frac{1}{2} (z_1, z_2) \widetilde{N_1} \Sigma N_2 \left( 
\begin{array}{c}
z_3 \\
z_4
\end{array} \right)\right] \nonumber \\
& & \mbox{\hspace{1cm}} \times
\exp \left[ (a^\dagger, a) N_1 \left( 
\begin{array}{c}
z_1 \\
z_2
\end{array} \right) +
(a^\dagger, a) N_1 \left( 
\begin{array}{c}
z_3 \\
z_4
\end{array} \right)
\right].
\eseq
\end{document}